\documentstyle[aps,prb,amsmath,amsfonts,amssymb,graphicx,multicol]
{revtex}
\newcommand{\nc}{\newcommand}
\nc{\sotimes}{\mathop{\otimes}_{s}}
\nc{\rnc}{\renewcommand}
\nc{\be}{\begin{equation}}
\nc{\ee}{\end{equation}}
\nc{\bea}{\begin{eqnarray}}
\nc{\eea}{\end{eqnarray}}
\nc{\bean}{\begin{eqnarray*}}
\nc{\eean}{\end{eqnarray*}}
\nc{\ba}{\begin{array}{@{\,}ll}}
\nc{\ea}{\end{array}}
\nc{\bc}{\begin{cases}}
\nc{\ec}{\end{cases}}
\nc{\lt}{\left\{}
\nc{\rt}{\right.}
\nc{\nn}{\nonumber}
\nc{\mb}{\mbox}
\nc{\Y}{{\cal Y}}
\nc{\T}{{\cal T}}
\nc{\st}{{\cal S}}
\nc{\tnh}{{\rm th}}
\nc{\sh}{{\rm sh}}
\nc{\ch}{{\rm ch}}
\nc{\mbbz}{{\mathbb Z}}
\nc{\odd}{\in2\mbbz+1}
\nc{\even}{\in2\mbbz}
\nc{\od}{\in2\mbbz+1}
\nc{\evn}{\in2\mbbz}
\nc{\img}{{\rm i}}

\newtheorem{cj}{Conjecture}
\begin{document}
%\sloppy
%\maketitle
%%%%%%%%%%%%%%%%%%%%%%%%%%%%%%%%%%%%%%%%%%%%%%%%%%%%%%%%%%%%%%%%%
\draft
\title{%
Excited state TBA and functional relations
in  spinless Fermion model
}
\author{%
Kazumitsu Sakai}
\address{%
Institute of Physics, University of Tokyo,
Komaba 3-8-1, Meguro-ku, Tokyo 153-8902, Japan}
\date{\today}
\maketitle
\begin{abstract}
The excited state thermodynamic Bethe ansatz 
(TBA) equations for the spinless Fermion model
are presented by the quantum transfer matrix 
(QTM) approach.
We introduce a more general family called
$T$-functions and explore
functional relations among them ($T$-system) 
and their certain combinations ($Y$-system).
{}From their analytical property,
we derive a closed set of non-linear integral
equations which characterize the correlation
length of $\langle c_j^{\dagger}c_i\rangle$
at any finite temperatures.
Solving these equations numerically, we explicitly
determine the correlation length, which coincides with
earlier results with high accuracy.
\end{abstract}
%%%%%%%%%%%%%%%%%%%%%%%%%%%%%%%%%%%%%%%%%%%%%%%%%%%%%%%%%%%%%%%%%
\begin{multicols}{2}
\narrowtext
%%%%%%%%%%%%%%%%%%%%%%%%%%%%%%%%%%%%%%%%%%%%%%%%%%%%%%%%%%%%%%%%
%%%%%%%%%%%%%%%%%%%%%%%%%%%%%%%%%%%%%%%%%%%%%%%%%%%%%%%%%%%%%%%%
%
Thermodynamics of 1D
quantum integrable systems
have been discussed by the thermodynamic Bethe ansatz
(TBA) method \cite{YY} based on the string hypothesis.
\cite{G,T,TS} 
Thermal quantities are determined 
by a set of non-linear integral equations
called TBA equations.
As an alternative method, 
the quantum transfer matrix
(QTM) approach has been proposed.
\cite{MSuzPB,InSuz,InSuz2,Koma,SAW,SNW,DdV,TakQT,Mizu,Klu,KZeit,JK,JKStJ,JKStp,JKSfusion,JKSHub,KSS,Klu2,FKM,FKM2,JSuz1,S3U}
It utilizes the Trotter formula
and reduces the calculation of the free energy to
the eigenvalue problem of the QTM on
a fictitious system.
A set of auxiliary functions, 
including the QTM itself,
satisfy functional relations.
Under suitable choice, these functions
have a property called ANZC (Analytic, 
NonZero and Constant asymptotics)
in  certain strips.
This admits a transformation of the 
functional relations to non-linear integral equations
(NLIE) which characterize the free energy.

Though each method reproduces the correct free energy,
there seems to be no relation between the QTM and the TBA.
Recently, however, we recognize the NLIE are identical to
the TBA equations by utilizing
the fusion hierarchy.
\cite{Klu,JKSfusion,KSS,JSuz1}
The members of fusion hierarchy ($T$-functions) 
and the combinations of them ($Y$-functions)
satisfy functional relations called $T$-system
and $Y$-system, respectively.(See ref.~\onlinecite{KNS1,KNS2}
for general relations between $T$ and $Y$-systems.)
By selecting these functions such that 
they are ANZC in appropriate strips, we  derive the
NLIE which are identical to the TBA equations.
Moreover considering the sub-leading eigenvalues, 
we derive systematically  
the ``excited state TBA" equations,
which is difficult within the string hypothesis. 

We apply this procedure to the spinless Fermion model.
The physical quantities such as the free energy and the 
correlation length have been already derived by 
more practical choice of the auxiliary functions
in ref.~\onlinecite{S3U}.
Our present interest, however, is 
the derivation of the TBA and
excited state TBA equations
by utilizing the fusion hierarchy and to confirm
the consistency between the resultant
equations and the earlier ones.
The Hamiltonian on the
periodic lattice of size $L$ is 
%\begin{eqnarray}
\[
{\cal H}=\frac{t}{2}\sum_{j=1}^{L} 
\Big\{ c_{j}^{\dagger}c_{j+1}\!+\!c_{j+1}^{\dagger} c_{j}\! 
%\nonumber \\
%& & \ \ \ \ \ 
+\!2\Delta\!\left(\!n_j\!-\!\frac{1}{2}\!\right)\! 
\left(\!n_{j+1}\! - \!\frac{1}{2} \!\right)\!  \Big\},
\label{eq.hamiltonian} 
%\end{eqnarray}
\]
where 
we consider the model in the repulsive critical region
$0\!\le\!\Delta\!=\!\cos\theta\!<\!1$, $0<t$ 
and introduce the parameter $p_0$ as
$p_0= \pi/\theta$\,($p_0\ge 2$). 
The commuting QTM on a fictitious
system of size $N$ (Trotter number, $N\in 2{\mathbb Z}$)
has been defined 
by the Fermionic $R$-operator and its super-transposition.\cite{S3U} 
The eigenvalue is expressed as 
\begin{eqnarray}
&&T_1(u,v)=\phi_{+}(v) \phi_{-}(v - 2\img)
    \frac{Q(v + 2\img)}{Q(v)} \nn \\
&&    \qquad \qquad \qquad+\varepsilon
%(-1)^{\frac{N}{2} + N_{\rm e}} 
     \phi_{-}(v) \phi_{+}(v  + 2\img) \frac{Q(v - 2\img)}
                  {Q(v)},  \\
&&\phi_{\pm}(v) = \left( 
     \frac{\sh \frac{\theta}{2} 
     (v \pm \img u)}{\sin\theta} \right)^{\frac{N}{2}}\!, \, 
     Q(v) = \prod_{j=1}^{N_{\rm e}} \sh \frac{\theta}{2} (v- w_j). \nn 
\eea
Here $\varepsilon\!=\!(-1)^{N/2+N_{\rm e}}\!$ and
$N_{\rm e}=\{0,1,\cdots,N/2 \}$ 
is the quantum number counting
the holes on odd sites and particles on even sites.
$\{w_j\}$ is a set of solutions to the Bethe ansatz equation (BAE):
\be
\left( \frac{\phi_{+}(v) \phi_{-}(v - 2\img)}
{\phi_{-}(v) \phi_{+}(v + 2\img )} \right)^{\frac{N}{2}}
=-\varepsilon\prod_{k =1}^{N_{\rm e}}
\frac{Q(w_j - 2\img)}{Q(w_j + 2\img)}. 
\label{eq.bae}
\end{equation} 
The largest eigenvalue $T_1^{(1)}(u,v)$ and the second largest
eigenvalue $T_1^{(2)}(u,v)$ lie in 
$N_{\rm e}\!=\!N/2$ and $N_{\rm e}\!=\!\!N/2\!-\!1$, respectively.
Hereafter we often omit the $u$ variable in $T_1(u,v)$.
The free energy per site $f$
is represented by
\be
f=-\frac{1}{\beta}\lim_{N\to\infty}\ln T_1^{(1)}(u_N,0)
-\frac{t}{4}\Delta,\,\,
u_N=-\frac{\beta t\sin\theta}{\theta N}.
\end{equation}
The correlation length $\xi$ of 
$\langle c_j^{\dagger}c_i\rangle$ can be also given
by the ratio of $T^{(1)}_1(v)$ to $T^{(2)}_1(v)$,
\be
\frac{1}{\xi}=
-\lim_{N\to\infty}\ln\Biggl|\frac{T_1^{(2)}(u_N,0)}{T_1^{(1)}(u_N,0)}\Biggr|.
\label{cl}
\end{equation}
To evaluate $T_1(v)$,
we consider a more general family called $T$-functions:
\bea
T_{n-1}(u,v)&&=\sum_{j=1}^n
     \varepsilon^{j-1}\phi_{-}\bigl(v-\img(n+2-2j)\bigr)\nn \\ 
     \times&&\frac{\phi_{+}\bigl(v-\img(n-2j)\bigr)Q(v+\img n)Q(v-\img n)}
    {Q\bigl(v+\img(2j-n)\bigr)Q\bigl(v+\img(2j-n-2)\bigr)},
\label{dvf}
\eea
where we set $T_{-1}(x)=0$.

For any $v \in \mathbb{C}$ and integers $n \ge m \ge 1$,
the $T$-functions satisfy the following relations,
\bea
T_{n-1}(v+\img m)T&&_{n-1}(v-\img m)= 
T_{n+m-1}(v)T_{n-m-1}(v) \nn \\
&&+\varepsilon^{n-m}T_{m-1}(v+\img n)T_{m-1}(v-\img n).
\label{tsystem1}
\eea
From now on we consider the case $p_0\in\mbbz_{\ge 3}$.
In this case $T$-functions satisfy further functional
relation,
\be
T_{p_0}(v) = 
\varepsilon T_{p_0-2}(v)+ 
(-1)^{N_{\rm e}}(1+\varepsilon^{p_0})T_{0}(v+\img p_0).
\label{tsystem2}
\end{equation}
We call the above two functional relations the $T$-system.
The proof of them are direct 
by using (\ref{dvf}) and the periodicity of the $T$-functions
\be
T_{n-1}(v)=T_{n-1}(v+2p_0 \img).
\end{equation}
Let us define a set of combinations among $T$-functions as
\bea
&&Y_{j}(v)=\frac{T_{j+1}(v)T_{j-1}(v)}{T_0(v+\img(j+1))T_0(v-\img(j+1))},
       \nn \\
&&\varepsilon^j+Y_{j}(v)=\frac{T_{j}(v+\img)T_{j}(v-\img)}
             {T_0(v+\img(j+1))T_0(v-\img(j+1))},\label{yfnc} \\
&&K(v)=(-1)^{N_{\rm e}} \frac{T_{p_0-2}(v)}{T_0(v+\img p_0)}.\nn 
\eea
where $1\le j\le p_0-2$, $Y_0{(v)}=0$ and $Y_{-1}(v)=\infty$.
We find that $\{Y_j(v) \}_{j=1}^{p_0 - 1}$ and $K(v)$
satisfy  following finite set of functional relations,
which we call the $Y$-system.
\bea
&&Y_{j}(v+\img)Y_{j}(v-\img)=(\varepsilon^{j-1}+Y_{j-1}(v))
                       (\varepsilon^{j+1}+Y_{j+1}(v)),\nn \\
&&\varepsilon^{p_0-1}+Y_{p_0-1}(v)=(\varepsilon+K(v))
                      (\varepsilon^{p_0}+\varepsilon K(v)),\label{ysystem} \\
&&K(v+\img)K(v-\img)=\varepsilon^{p_0}+Y_{p_0-2}(v),\nn
\eea
This can be proved by the $T$-system (\ref{tsystem1}) and
(\ref{tsystem2}).
Let $T^{(k)}_{n-1}(v)$ denote $T_{n-1}(v)$
constructed by the BAE roots which is relevant
to the $k$-th largest eigenvalue 
and $Y_j^{(k)}(v)$, $K^{(k)}(v)$ be the $Y$-functions
constructed from $\{T_{n-1}^{(k)}\}$ as in (\ref{yfnc}).
Using the property
$\phi_{\pm}(v\pm 2p_0 \img)=(-1)^{N/2}\phi_{\pm}(v)$ and
$Q(v+2p_0 \img)=(-1)^{N_{\rm e}/2}Q(v)$,  
we can easily show for $p_0\in 2{\mathbb Z}$
\be
T_{p_0-1}^{(2)}(v)=0, \quad \,\, Y_{p_0-2}^{(2)}(v)=0 \label{t0}.
\end{equation}

To proceed further, one clarifies the analyticity of the $Y$-functions
by numerical studies on the
$T$-functions, keeping the Trotter number
$N$ finite.
{}From them, after suitable modification
we confirm the $Y$-functions have the ANZC 
property in the strip $\mb{Im}v\in[-1,1]$.
We call this  ``physical strip".
Then we can transform the $Y$-system to a closed set of NLIE
in the following way.
First we take the logarithmic derivative
and perform the Fourier transformation
on both side of the first and third equation
in (\ref{ysystem}).
Second by  Cauchy's theorem, 
the Fourier mode for the logarithmic derivative of $Y_j(v)$
($K(v)$) is expressed by those of $\varepsilon^{j+1}+Y_{j+1}(v)$
and $\varepsilon^{j-1}+Y_{j-1}(v)$ ($\varepsilon^{p_0}+
Y_{p_0-2}(v)$). 
Finally performing the inverse Fourier transformation
and integrating over $v$,
we derive the desired  NLIE.
In these NLIE, the Trotter limit $N\to\infty$ can be
taken analytically.
The eigenvalue $T_1^{(k)}(v)$ are determined after performing such
transformation on 
\be
T_1(v+i)T_1(v-i)=(\varepsilon+Y_1(v))T_0(v+2i)T_0(v-2i).
\label{tdetermine}
\end{equation}

For the largest eigenvalue in the sector $N_{\rm e}=N/2$,
all functions and functional relations 
are equivalent to those of the $XXZ$ model in
the critical regime.\cite{KSS}
Therefore one concludes that the resultant NLIE 
which characterize the free energy 
for the spinless fermion model
are identical to the TBA equations for the
$XXZ$ model.\cite{TS,KSS}

Let us consider the second largest eigenvalue in the 
sector $N_{\rm e}=N/2-1$.
As is mentioned in ref.~\onlinecite{S3U},
we find that two pure imaginary eigenvalues 
which are complex conjugate each other, 
are degenerate in magnitude.
All the BAE roots for these two eigenvalues 
are real.
The two distributions of the BAE roots
are symmetrical with respect to the imaginary axis.
Hereafter we consider the case $\mb{Im} T_1^{(2)}(0)>0$
(the case $\mb{Im} T_1^{(2)}(0)<0$ also can be discussed
in similar way).
According to the numerical studies, 
we have the following conjecture
which is quite different from that in $XXZ$ model.\cite{KSS}
\begin{cj}
$T^{(2)}_{n-1}(v)$ has one real zero $\zeta_{n-1}$
for $\{n\even|2\le n\le p_0-1\}$ and 
two real zeros $\zeta_{n-1}$ and $-\zeta^{\prime}_{n-1}$
for $\{n\odd|2< n< p_0/2\}$.
All other zeros are out of the physical strip.
\end{cj}
Here $\zeta_{n-1},\zeta_{n-1}^{\prime}>0$.
For example we depict the location of zeros
in Fig.~\ref{zeros} for $p_0=10$.
\begin{figure}
\begin{center}
\includegraphics[width=0.48\textwidth]{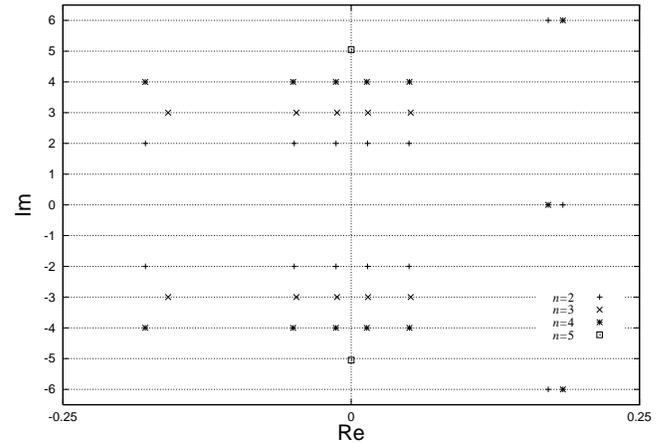}
\end{center}
\caption{Location of zeros for $T_{n-1}^{(2)}(u,v)$ for
$u=-0.05$, $p_0=6$, $N=12$.
There exists one real zero for $n=2$ and $n=4$, while
for $n=3$ two zeros are at $\pm \infty$.
Zeros on the imaginary axis for $n=5$ are sextuple roots.
}
\label{zeros}
\end{figure}
$Y^{(2)}_j$ and $K^{(2)}$ have the asymptotic values
\begin{subequations}
\bea
 Y^{(2)}_j &\rightarrow& 
         \lt
             \ba
              -\frac{\sin(j+2)\theta \sin j\theta}
                    {\cos^2 \theta} 
              & \mb{for $j\in 2\mbbz$} \\
               \frac{\cos(j+2)\theta\cos j \theta}
                    {\cos^2 \theta}
              & \mb{for $j\in 2\mbbz+1$}
             \ea \rt, \label{yasympt} \\
 K^{(2)} &\rightarrow&
          \lt
           \ba
             1 
             & \mb{for $p_0\in 2\mbbz$} \\
             \img \tan\theta \,\,(v\to\infty)
             & \mb{for $p_0\in 2\mbbz+1$} \\
             -\img\tan\theta \,\, (v\to -\infty)
             & \mb{for $p_0\in 2\mbbz+1$}
           \ea
     \rt.\label{kasympt}
\eea
\end{subequations}
{}From  conjecture 1 and above asymptotics,
the factors of the rhs in (\ref{ysystem}) 
such as $(-1)^{j+1}+Y^{(2)}_{j-1}(v)$ 
are ANZC in 
$\mb{Im} v \in [-\epsilon,\epsilon]$ ($\epsilon\ll 1$).
For the lhs, we need to modify 
$Y^{(2)}_j(v)$ ($\{j\even|2\le j\le p_0-3\}$ or
$\{j\odd|1\le j\le p_0/2\}$) and $K^{(2)}(v)$  
($p_0\odd$) such that 
they are ANZC in the physical strip. 
For example, we take the case $p_0=6$.
The following modification,
\bean
\widetilde{Y}_1^{(2)}(v)
    &=&\frac{Y_1^{(2)}(v)e^{\frac{\pi}{p_0}v\tnh\frac{\pi}{4}v}}
{\left(\tnh\frac{\pi}{4}(v-\img(u+1))\tnh\frac{\pi}{4}(v+\img(u+1))\right)^
           {\frac{N}{2}}},\nn \\
\widetilde{Y}_2^{(2)}(v)&=&\frac{Y_2^{(2)}(v)}
{\tnh\frac{\pi}{4}(v-\zeta_1)\tnh\frac{\pi}{4}(v-\zeta_3)},\nn \\
\widetilde{Y}_3^{(2)}(v)&=&Y_3^{(2)}(v)e^{\frac{\pi}{p_0}v\tnh\frac{\pi}{4}v}, 
\eean
is sufficient.
Note that the factor $e^{\pi v\tnh(\pi v/4)/p_0}$ is included
to compensate the singularity caused by $Y^{(2)}_1(v)$ and
$Y^{(2)}_3(v)$ tending to zero as $e^{-\pi|v|/p_0}$ at $v=\pm\infty$.
{}Let us rewrite the  $Y$-functions in the Trotter limit as
\be
\eta_j(v)\!=\!\lim_{N\to\infty}Y^{(2)}_j(u_N,v),\,
\kappa(v)\!=\!\lim_{N\to\infty}K^{(2)}(u_N,v).\!\!\!\!
\label{yklimit}
\end{equation}
After taking the logarithmic derivative and
performing the  Fourier transformation,
we derive the NLIE obeyed
by $\eta_j$ and $\kappa$.
%
%\begin{full}
\end{multicols}
\widetext
\noindent
\setlength{\unitlength}{1in}
\begin{picture}(0.375,0)
  \put(0,0){\line(1,0){3.375}}
  \put(3.375,0){\line(0,1){0.08}}
\end{picture}

\noindent
\bea
\ln &&\eta_1(v)= 
-\frac{\beta\pi t \sin\theta}{2\theta\ch(\frac{\pi v}{2})} 
+s_1\ast\ln((1+\eta_2)h_1)(v) 
+    \lt \ba
          \pi \img
              & \mb{for $p_0\le 5$ } \\
          \pi \img-\frac{\pi}{6}v\tnh\frac{\pi v}{4}
              & \mb{for $p_0=6$} \\
           \ln\{
               \tnh\frac{\pi}{4}(v-\zeta_2)
              \tnh\frac{\pi}{4}(v+\zeta_2^{\prime})
              \}     
              & \mb{for $p_0 \ge 7$  }
        \ea \rt, \nn \\
\ln &&\eta_j(v)=
    s_1\ast\ln(1-\eta_{j-1})(1-\eta_{j+1})(v) 
    + \ln \{ \tnh\frac{\pi}{4}(v-\zeta_{j-1})
    \tnh\frac{\pi}{4}(v-\zeta_{j+1}) \}+\pi \img \,\,\,\,
\mbox{for } \{j\even|2\le j\le p_0-3\}, \nn \\ 
\ln && \eta_j(v)= 
s_1\ast\ln((1+\eta_{j-1})(1+\eta_{j+1})h_j)(v)\nn \\
&&\hspace{1cm}+
    \lt  \ba
          \ln \{\tnh\frac{\pi}{4}(v-\zeta_{j-1})
                 \tnh\frac{\pi}{4}(v+\zeta_{j-1}^{\prime}) 
                 \tnh\frac{\pi}{4}(v-\zeta_{j+1})
                 \tnh\frac{\pi}{4}(v+\zeta_{j+1}^{\prime})
                 \}
            & \mb{for $\{j\odd|3\le j <\frac{p_0-4}{2}\}$} \\
       \ln\{\tnh\frac{\pi}{4}(v-\zeta_{j-1})
                 \tnh\frac{\pi}{4}(v+\zeta_{j-1}^{\prime})
                 \} +\pi \img-\frac{\pi}{p_0}v\tnh\frac{\pi}{4} v
            & \mb{for $\{j\odd|j=\frac{p_0-4}{2}\}$} \\
       \ln\{\tnh\frac{\pi}{4}(v-\zeta_{j-1})
                 \tnh\frac{\pi}{4}(v+\zeta_{j-1}^{\prime})
                 \} 
                 +\pi \img
         & \mb{for $\{j\odd|\frac{p_0-4}{2}<j<\frac{p_0}{2}\}$} \\
          -\frac{\pi}{p_0}v\tnh\frac{\pi}{4} v
          & \mb{for $\{j\odd|j=\frac{p_0}{2}\}$} \\
          0
          & \mb{for $\{j\odd|\frac{p_0}{2}<j \le p_0-3\}$}
      \ea \rt, \label{nlie} \\
 &&\lt \ba
     \eta_{p_0-2}(v)=0,\,\,\,\,\kappa(v)=1 & \mb{for $p_0\evn$} \\
     \eta_2(v)=-\kappa^2(v) &\mb{for $p_0=3$} \\
        \ln\eta_{p_0-2}(v)=
           s_1\ast\ln(1+\eta_{p_0-3})(1-\kappa^{2})(v)
                               & \mb{for $p_0\od$ and $p_0\ne 3$} \\
        \ln\kappa(v)=
           s_1\ast\ln(1-\eta_{p_0-2})(v)+
           \ln\{\tnh\frac{\pi}{4}(v-\zeta_{p_0-2})\}+\frac{\pi}{2} \img
                                     &\mb{for $p_0\od$ }
  \ea \rt \nn,           
\eea
%\end{full}
\noindent
\hfill
\begin{picture}(3.375,0)
  \put(0,0){\line(1,0){3.375}}
  \put(0,0){\line(0,-1){0.08}}
\end{picture}
\begin{multicols}{2}
\narrowtext
\noindent
where
\[
 h_j(v)\!=\!\lt \ba
           \!\exp\!
             \left(\!\frac{2\pi}{p_0}\!\left(\!\frac{v \sh\frac{\pi}{2}v-1}
             {\ch\frac{\pi}{2}v}\!\right)\!\right)\!\! 
              &\!\mb{for\,$\{j\in 2{\mathbb Z}\!+\!1|
              j\!=\!\frac{p_0-4}{2}\! ,
                  \!\frac{p_0}{2}\}$}\!\!\!
              \\
              1 \!\!&\mb{otherwise}.
         \ea \rt
\]
The symbol $\ast$ denotes the convolution
\[
f\ast g(x)=\int_{-\infty}^{\infty}f(x-y)g(y)dy,
\]
and
$s_1(v)=1/4\ch\frac{\pi}{2}v$.
Here the integration constants have been fixed from the asymptotic
values (\ref{yasympt}) and (\ref{kasympt}).
In addition to above equations, we need to impose the 
consistency condition coming from $T_j^{(2)}(\zeta_j)=0$
and  $T_j^{(2)}(-\zeta_j^{\prime})=0$.
{}From (\ref{yfnc}) and (\ref{yklimit}), this leads to 
$\eta_j(\zeta_j\pm \img)=1$ for $\{j\odd|1\le j \le p_0-2\}$ and
$\eta_j(\zeta_j\pm \img)=\eta_j(\zeta_j^{\prime}\pm \img)=-1$ 
for $\{j\even|1< j <p_0/2-1\}$.
Explicitly they read
%
%\begin{full}
\end{multicols}
\widetext
\noindent
\setlength{\unitlength}{1in}
\begin{picture}(3.375,0)
  \put(0,0){\line(1,0){3.375}}
  \put(3.375,0){\line(0,1){0.08}}
\end{picture}
\noindent
\bea
0&&= 
i\frac{\beta\pi t \sin\theta}{2\theta\ch(\frac{\pi \zeta_1}{2})} 
+s_1\ast\ln((1+\eta_2)h_1)(\zeta_1+i) 
+    \lt \ba
        -\pi i
              &\mb{for $p_0\le 5$} \\
          -\pi i-\frac{\pi}{6}(\zeta_1+\img)
              \tnh\frac{\pi (\zeta_1+\img)}{4}
              & \mb{for $p_0=6$}     \\
           \ln\{
               \tnh\frac{\pi}{4}(\zeta_1-\zeta_2+\img)\}+
           \ln\{\tnh\frac{\pi}{4}(\zeta_1+\zeta_2^{\prime}+\img)
              \}-2\pi \img     
              & \mb{for $p_0 \ge 7$}
      \ea \rt, \label{z1} \nn \\
0&&= \lt \ba
    s_1\ast\ln(1-\eta_{j-1})(1-\eta_{j+1})
    (\zeta_j+\img) 
    + \ln \{ \tnh\frac{\pi}{4}(\zeta_j-\zeta_{j-1}+\img)\}+
    \ln\{\tnh\frac{\pi}{4}(\zeta_j-\zeta_{j+1}+\img) \}  \\
    s_1\ast\ln(1-\eta_{j-1})(1-\eta_{j+1})(\zeta_j^{\prime}+\img)
    + \ln \{ \tnh\frac{\pi}{4}(\zeta_j^{\prime}-\zeta_{j-1}+\img)\}+ 
    \ln\{\tnh\frac{\pi}{4}(\zeta_j^{\prime}-\zeta_{j+1}+\img) \}
    -2\pi \img  
       \ea \rt\nn \\
&& \hspace{12cm}
\mbox{for $\{j\even|2\le j < \frac{p_0}{2}-1\}$}, \nn \\
0&&= 
s_1\ast\ln((1+\eta_{j-1})(1+\eta_{j+1})h_j)(\zeta_j+\img) \label{nliez} \\
+&&\lt
     \ba 
        \ln \{\tnh\frac{\pi}{4}(\zeta_j\!-\zeta_{j-1}+\img)\}+
          \ln \{\tnh\frac{\pi}{4}(\zeta_j\!+
                 \zeta_{j-1}^{\prime}+\img)\} \\ 
         +\ln\{\tnh\frac{\pi}{4}(\zeta_j\!-\zeta_{j+1}+\img)\}+
          \ln\{\tnh\frac{\pi}{4}(\zeta_j\!+\zeta_{j+1}^{\prime}+\img)
                 \}-2\pi \img 
            & \mb{for $\{j\odd|3\le j <\frac{p_0-4}{2}\}$} \\
       \ln\{\tnh\frac{\pi}{4}(\zeta_j\!-\zeta_{j-1}+\img)\}+
       \ln\{\tnh\frac{\pi}{4}(\zeta_j\!+\zeta_{j-1}^{\prime}+\img)
                 \} 
       -\pi \img-\frac{\pi}{p_0}(\zeta_j\!+\img)\tnh\frac{\pi}{4} 
                (\zeta_j\!+\img)
            &\mb{for $\{j\odd|j=\frac{p_0-4}{2}\}$} \\
       \ln\{\tnh\frac{\pi}{4}(\zeta_j\!-\zeta_{j-1}+\img)\}+
       \ln\{\tnh\frac{\pi}{4}(\zeta_j\!+\zeta_{j-1}^{\prime}+\img)
                 \}-\pi \img
         & \mb{for $\{j\odd|\frac{p_0-4}{2}<j<\frac{p_0}{2}\}$} \\
           -\frac{\pi}{p_0}(\zeta_j\!+\img)\tnh\frac{\pi}{4} 
          (\zeta_j+\img)
          & \mb{for $\{j\odd|j=\frac{p_0}{2}\}$} \\
          0
          & \mb{for $\{j\odd|\frac{p_0}{2}<j \le p_0-3\}$} \nn 
      \ea
  \rt, \nn \\ 
0&&=s_1\ast\ln(1+\eta_{p_0-3})(1-\kappa^{2})(\zeta_{p_0-2}+\img)
 \hspace{6.1cm}  \mbox{for $p_0\od$  and $p\ne 3$} \nn,
\eea
%\end{full}
%\noindent
%\hfill
%\begin{picture}(3.375,0)
%  \put(0,0){\line(1,0){3.375}}
%  \put(0,0){\line(0,-1){0.08}}
%\end{picture}
%%
\begin{multicols}{2}
\narrowtext
\noindent
where the convolutions should be interpreted as
\[
 s_1*g(\zeta+\img)=\mbox{p.v.}\left(\int^{\infty}_{-\infty}\frac{g(x)}
{4\img\sh\frac{\pi}{2}(\zeta-x)}dx\right)+\frac{1}{2}g(\zeta).
\]
Here p.v. means the principal value.
Then the Trotter limit of $T_1^{(2)}(v)$ can be expressed from
(\ref{tdetermine}) as
\bea
 \lim_{N\to\infty}&& T^{(2)}_1(u_N,0)
     =\ln\tnh\frac{\pi}{4}\zeta_1 \nn \\ 
   &&+\int^{\infty}_{-\infty}dv s_1(v)\ln(1-\eta_1(v)) \nn \\
   &&+t\beta\int^{\infty}_{-\infty}dv s_1(v)\frac{\sin^2\theta}
      {\ch\theta v-\cos\theta} 
      -\frac{t}{2}\beta\cos\theta+\frac{\pi}{2}\img. \nn 
 \eea
Especially for $p_0=4$, the zero $\zeta_1$ and
$Y$-functions $\eta_1(v)$ are explicitly determined
by using the fact (\ref{t0}), which is a novel feature in
the present approach.\cite{S3U}
\[
\eta_1(v)=-\exp\left(-\frac{\sqrt{2} \beta t}
{\ch\left(\frac{\pi}{2} v \right)}\right),\quad
\zeta_1=\frac{2}{\pi}\ch^{-1}\left(\frac{\sqrt{2}
\beta t}{\pi}\right). 
\]
Finally we obtain the correlation length (\ref{cl}) as 
\be
\frac{1}{\xi}=-\ln\tnh\frac{\pi}{4}\zeta_1-
               \int^{\infty}_{-\infty}dv s_1(v)\ln\left
               (\frac{1-\eta_1(v)}{1+\eta^{(1 )}_1(v)}\right),
\label{correlation}
\end{equation}
where the function $\eta_1^{(1)}(v)$ characterizing
the free energy has been calculated in
ref.~\onlinecite{TS,KSS}.
We evaluate (\ref{nlie}) and (\ref{nliez}) numerically and explicitly
determine the correlation length in Fig.~\ref{sfmcl}.
\begin{figure}
\begin{center}
\includegraphics[width=0.48\textwidth]{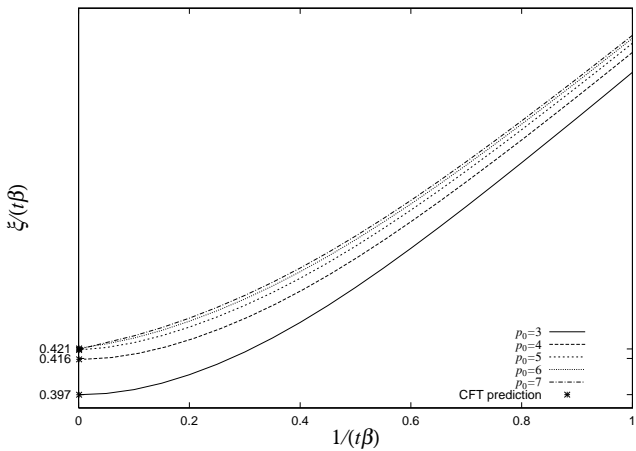}
\end{center}
\caption{Ratio of the correlation length and
inverse temperature for $p_0=3,4,5,6$ and 7.}
\label{sfmcl}
\end{figure}
These results are
consistent with those in ref.~\onlinecite{S3U}. 
Especially in the low temperature limit they
agree with known expression from CFT 
\be
 \lim_{\beta\to\infty}
\xi_2(\beta)/\beta=\frac{t\sin \theta}{2\theta}
                   \left(\frac{\pi-\theta}{\pi}+
                         \frac{\pi}{4(\pi-\theta)}
                   \right)^{-1}.
\label{known1}
\end{equation}
\acknowledgements
The author is grateful to A. Kuniba and J. Suzuki 
for useful discussions and critical reading of the manuscript.
He also thanks G. Hatayama, M. Shiroishi and 
Y. Umeno for discussions and helpful comments.
%
%%%%%%%%%%%%%%%%%%%%%%%%%%%%%%%%%%%%%%%%%%%%%%%%%%%%%%%

\end{multicols}
\end{document}